\documentclass[aps,showpacs,preprintnumbers]{revtex4}
\usepackage{graphicx}
\usepackage{amsmath}
\usepackage{amsfonts}
\usepackage{amssymb}

\begin{document}
\title{
Reply to "Comment on equivalence between quantum phase transition
phenomena in radiation-matter and magnetic systems"}
\author{Giuseppe Liberti} \email{liberti@fis.unical.it}
\author{Rosa Letizia Zaffino}
\affiliation{
Dipartimento di Fisica, Universit\`a della Calabria\\
INFN - Gruppo collegato di Cosenza, 87036 Rende (CS) Italy}
\date{\today}
 \maketitle
In their Comment \cite{brankov}, Brankov, Tonchev and Zagrebnov,
claim that the temperature-dependent effective Hamiltonian derived
in Ref. \cite{libzaf} from a radiation-matter Dicke model violate
a "rigorous" result that the same authors have obtained 30 years
ago \cite{brankov2}. It is clear that Brankov, Tonchev and
Zagrebnov have misunderstood the results of Ref. \cite{libzaf} in
several ways.
\begin{enumerate}
    \item First, the temperature-dependent effective Hamiltonian
given by
\begin{equation}\label{1}
    H(\beta)=\omega\, a^\dagger
   a+\frac{\epsilon}{2}S_z-\frac{\beta\lambda^2}{2N}\coth{\left(\frac{\beta\omega}{2}\right)}S_x^2
\end{equation}
is correct for $\beta^3\lambda^2\omega<1$ and $\beta\epsilon\ll
1$, i.e. only in the high-temperature limit. It is not permissable
to take the limit $\beta\rightarrow\infty$ to infer critical
properties and some connection with the collective one-dimensional
Ising model. For $\beta \rightarrow 0$ this Hamiltonian obviously
becomes
\begin{equation}\label{2}
    H=\omega\, a^\dagger
   a+\frac{\epsilon}{2}S_z-\frac{\lambda^2}{N\omega}S_x^2
\end{equation}
    \item  We have used the Zassenhaus formula \cite{wilcox} in its simplest
    form, through a perturbative expansion of partition function obtained by
decomposing the Hamiltonian into two non-commuting hermitian
operators. The results of our analysis are used to show an obvious
similarity with the result of Ref. \cite{reslen} but only in the
high-temperature limit, keeping the lowest-order terms of the
Polatsek and Becker expansion \cite{approx3}. It is possible to
derive higher-order approximations in a systematic manner but the
increasing complexity of the expressions requires numerical
calculations in order to derive the thermodynamical properties of
the system.
    \item We use the term "classical" with respect to the limit
    reached when $\beta\omega\ll 1$ where the results are exactly
    those obtained with the Wang and Hioe computational method.
    \cite{WH}
\end{enumerate}
Finally, we want to emphasize that the equivalence of the Dicke
Model and the Ising Model at equilibrium was originally discussed
by R. W. Gibberd \cite{gibberd}, who applied a Bogoliubov
transformation \cite{bogo} (the so-called "thermodynamically
equivalent Hamiltonian" method) to the field part of the Dicke
Hamiltonian.

    %%%%%%%%%%%%%%%%%%%%%%%%%%%%%%%%%%%%%%%%%%%%%%%%%%%%%%%%%%%%%%%%%

\end{document}